\documentclass[12pt,a4paper]{article} 
\usepackage{epsfig}
\usepackage{graphics}
\usepackage{subfigure}
\usepackage{a4wide}
\usepackage{amsmath}

\title{Planar Skyrmions at High and Low Density}
 \author{R.\ S.\ Ward\footnote{email: richard.ward@durham.ac.uk}
 \bigskip
\\Department of Mathematical Sciences,  \\ University of
Durham, \\Durham DH1 3LE}

\newcommand{\half}{{\scriptstyle\frac{1}{2}}}

\newcommand{\RR}{{\bf R}}

\newcommand{\vphi}{\vec\phi}
\newcommand{\pa}{\partial}

\newcommand{\ii}{{\rm i}}

\begin{document}
\maketitle
\abstract{\noindent
The O(3) Skyrme system in two space dimensions admits topological
soliton solutions.  This paper studies the transition between
the high-density crystalline phase of such solitons and the low-density
phase where there are multi-Skyrmions localized in space.  The details
depend crucially on the choice of the potential function.  Two such choices
are investigated: in the first system, multi-Skyrmions at low density
form a ring; while in the second, there is an explicit crystal solution,
and the preferred low-density configurations are chunks of this crystal.
The second system is a particularly good analogue of three-dimensional
Skyrmions.
}
\newpage

\section{Introduction}

Topological solitons are of interest both mathematically and in
many areas of physics (as models for particles, topological defects in
condensed-matter physics, etc).  A particularly important question is the
nature of, and transition between, the high-density and low-density
phases of such solitons.  Ultimately, one would like to understand
the thermodynamics of these systems; the starting-point for this
is to investigate their zero-temperature behaviour.  At its most basic,
this amounts to considering a fixed number $N$ of solitons confined
to a finite volume, and investigating the way in which static
classical $N$-soliton solutions depend on the volume (or density).
Alternatively, one may use the dimensionless ratio
{\em (size of soliton)/(size of space)} as a parameter.

The typical situation is as follows.  At large density, there is a
high degree of symmetry and uniformity, and in particular it may not
possible to identify individual solitons; in Skyrme models, for example,
one gets a periodic crystal-like structure of half-Skyrmions.
At low densities, by contrast, solitons (or multi-solitons) become
localized in space, and there is less symmetry.  The localized
multi-solitons may be `chunks' of the high-density crystal, or may
have a quite different (for example shell-like) shape --- this depends
on the details of the system.

The picture for the three-dimensional Skyrme model may be
summarized as follows.  Let $N$ denote the Skyrme number, and $E$
the normalized energy-per-Skyrmion (so the Bogomolny-Faddeev
bound is $E\geq1$).  At high density, the ground state is a
triply-periodic lattice of half-Skyrmions \cite{K85,GM87,KS88,W88,KS89,
CJJVJ89}, with energy $E=1.038$ at its most favourable density.
At low density (for example for an isolated multi-Skyrmion in $\RR^3$),
and for relatively low values of $N$ (in particular for $N\leq22$),
the minimal-energy static Skyrmions take the form of polyhedral shells
\cite{BS02}.  Their normalized energy $E$ is a decreasing function of $N$,
and it appears \cite{BS98} that these polyhedra have energy $E\approx1.06$
for large $N$.  Since this is larger than the energy of the Skyrme
crystal, one expects that there is a critical value $N_{{\rm c}}$ such
that for $N>N_{{\rm c}}$, the minimal-energy Skyrmion resembles a chunk
of the Skyrme crystal.  Such a lattice chunk will have energy
$E \approx 1.038 + kN^{-1/3}$, with the second term being a surface
contribution.  There have been attempts \cite{B96} to estimate the
constant $k$, and hence to determine $N_{{\rm c}}$, but these have
not been definitive.  There are also other possibilities for the
shape of large-$N$ Skyrmions, for example a multi-shell structure
\cite{MP01}.

The present paper deals with the two-dimensional O(3) Skyrme system, which one
may view as an analogue of the three-dimensional case, as well as being
of interest in its own right.  The 2-D Skyrme system contains, as a
limiting case, the two-dimensional O(3) sigma model; the sigma-model
crystal, obtained by imposing periodic boundary conditions, has been
studied both as a model quantum field theory \cite{RR83} and in
connection with the dynamics of (classical) solitons \cite{CZ97,S98}.
But sigma-model solitons do not have a fixed size, and in particular
tend to shrink and decay; so in that sense they are not true solitons.
We shall consider only the case where both the Skyrme term and a potential
term $V$ are present, and the soliton size is consequently fixed.
The interest here is in the nature of the low-density and the high-density
configurations, and the transition between them; as we shall see,
the details of these depend crucially on the choice of potential $V$.
Two different systems, corresponding to two different choices of $V$,
will be investigated in detail.


\section{Two-Dimensional Skyrmions}

The two-dimensional Skyrme system involves a unit 3-vector field
$\vphi=(\phi_1,\phi_2,\phi_3)$
defined on (2+1)-dimensional space-time.  We are interested
only in static configurations, so $\vphi$ depends on the spatial
variables $x^j=(x^1,x^2)$, thought of as (local) coordinates on
a two-dimensional space $S$.  If $S$ is a compact surface, then $\vphi$
has a winding number $N$, which we think of as the number of Skyrmions.
One special case is where space is the plane $\RR^2$ with boundary
condition $\vphi\to(0,0,1)$ as $r\to\infty$ --- this corresponds
(by conformal invariance)
to the 2-sphere $S=S^2$.  Another case is that of periodic boundary
conditions $\vphi(x^1,x^2)=\vphi(x^1+L,x^2)=\vphi(x^1,x^2+L)$,
which corresponds to the 2-torus $S=T^2$.  In either of these cases,
we may regard the spatial metric as being the flat (Euclidean) metric,
and we shall do so in what follows.

The integer $N$ can be either positive or negative; without loss of
generality, we shall assume $N$ to be positive.
The energy density of a configuration $\vphi$ is
\begin{equation}\label{Enden}
 {\cal E} = \half(\pa_j\vphi)\cdot(\pa_j\vphi)
             + \half\alpha\Omega^2 + \half\alpha V(\vphi),
\end{equation}
where $\Omega = \vphi\cdot\pa_1\vphi\times\pa_2\vphi$,
$V$ is some potential function, and $\alpha$ is a dimensionless constant.
The length-scale in this system is determined by the ratio of the
coefficients of the $\Omega^2$ and $V$ terms, and so is fixed
by choosing these coefficients to be equal as in (\ref{Enden}).
Static multi-Skyrmion solutions are critical points of the
(normalized) energy functional
\begin{equation}\label{En}
   E = \frac{1}{4\pi N} \int {\cal E} \,d^2x.
\end{equation}
There is a topological (Bogomolny) lower bound on the energy, namely
\begin{equation}\label{Bogbound}
   E \geq 1 + \frac{\alpha}{4\pi} \int_{S^2} \sqrt{V(\vphi)} \,d\omega,
\end{equation}
where $d\omega$ is the usual area element on the space $S^2$ of unit
vectors (in other words, $\int d\omega = 4\pi$).  Under certain
circumstances,
this bound can be saturated; the following statement is adapted from
\cite{IRPZ92}, see also \cite{IW01}. Write $z=x^1+\ii x^2$, and let
$W=(\phi_1+\ii\phi_2)/(1-\phi_3)$ denote the stereographic projection
of $\vphi$. Let $W(z)$ be a complex-analytic function satisfying a
first-order ordinary differential equation of the form $dW/dz=F(W)$
for some function $F$.  Then the configuration $W(z)$ saturates the
bound (\ref{Bogbound}), and hence is a static Skyrmion solution, provided
that the potential $V$ is given by
\begin{equation}\label{Bogpot}
  V = \frac{16\,|F(W)|^4}{(1+|W|^2)^4}.
\end{equation}
The simplest example of this is where $F$ is a constant; the corresponding
system (on~$\RR^2$) has been investigated in some detail, for example in
\cite{LPZ90b,Sut91}. In this case, there is a repulsive force
between Skyrmions, and consequently there are no static multi-Skyrmion
solutions.

For any system of the form (\ref{Enden}), the nature and shape of multi-Skyrmion
solutions depend on the choice of $V$, and many different choices have
been considered \cite{IRPZ92,ESZ00}.  The following two sections will
deal with two possible choices.  They are both analogous to the
three-dimensional Skyrme system, in that they allow both crystal-like
and ring-like solutions (the latter being the counterpart of the polyhedral
shells mentioned earlier).

To conclude this section, it is worth mentioning yet another
case which has been extensively investigated, namely
$V(\vphi) = 1-\phi_3$ \cite{PSZ95a,PSZ95b}.  In this case, the static
multi-Skyrmion solutions on $\RR^2$ appear to form a lattice-like structure
\cite{PSZ95a,ESZ00} --- for example, for even values of $N$ one gets
a lattice of double-Skyrmions as the lowest-energy state. There
are also other local minima of the energy, but there do not appear to
be any ring-like solutions (except for $N=2$).  The thermodynamics
of this system has been studied numerically \cite{SW02}.


\section{The System $V = 1-\phi_3^2$}

In this section, the potential function $V$ appearing in (\ref{Enden})
is taken to be $V(\vphi) = 1-\phi_3^2$.
Let us discuss, first, the localization-delocalization transition for
this system.  A useful order parameter in this regard is the quantity
\[
  \langle\phi_3\rangle = \frac{1}{A} \int_S \phi_3\,d^2x,
\]
where $A$ is the area of the 2-space $S$.  At high density, there is a
high degree of symmetry, and we expect to have $\langle\phi_3\rangle=0$;
while at low density, the field will localize in space and we will have 
$\langle\phi_3\rangle\neq0$.  The transition between these two phases
occurs at some critical density $\rho_{{\rm c}}$.
This transition was investigated by taking space to be a sphere
($S=S^2$) in \cite{IW01}; in particular, this paper looked at
$\langle\phi_3\rangle$, as a function of $A$, for a single skyrmion on
$S^2$.  Using approximate analytic expressions for the Skyrmion indicated
that the transition occurs at $A=4\pi\sqrt{5}$, {\em ie} the critical density
is
\begin{equation}\label{NBScrit1}
  \rho_{{\rm c}}=1/(4\pi\sqrt{5})=0.036\,.
\end{equation}
Numerical simulations gave results that supported this estimate.

\begin{figure}[htb]
\begin{center}
{
\subfigure{
\includegraphics[scale=0.4]{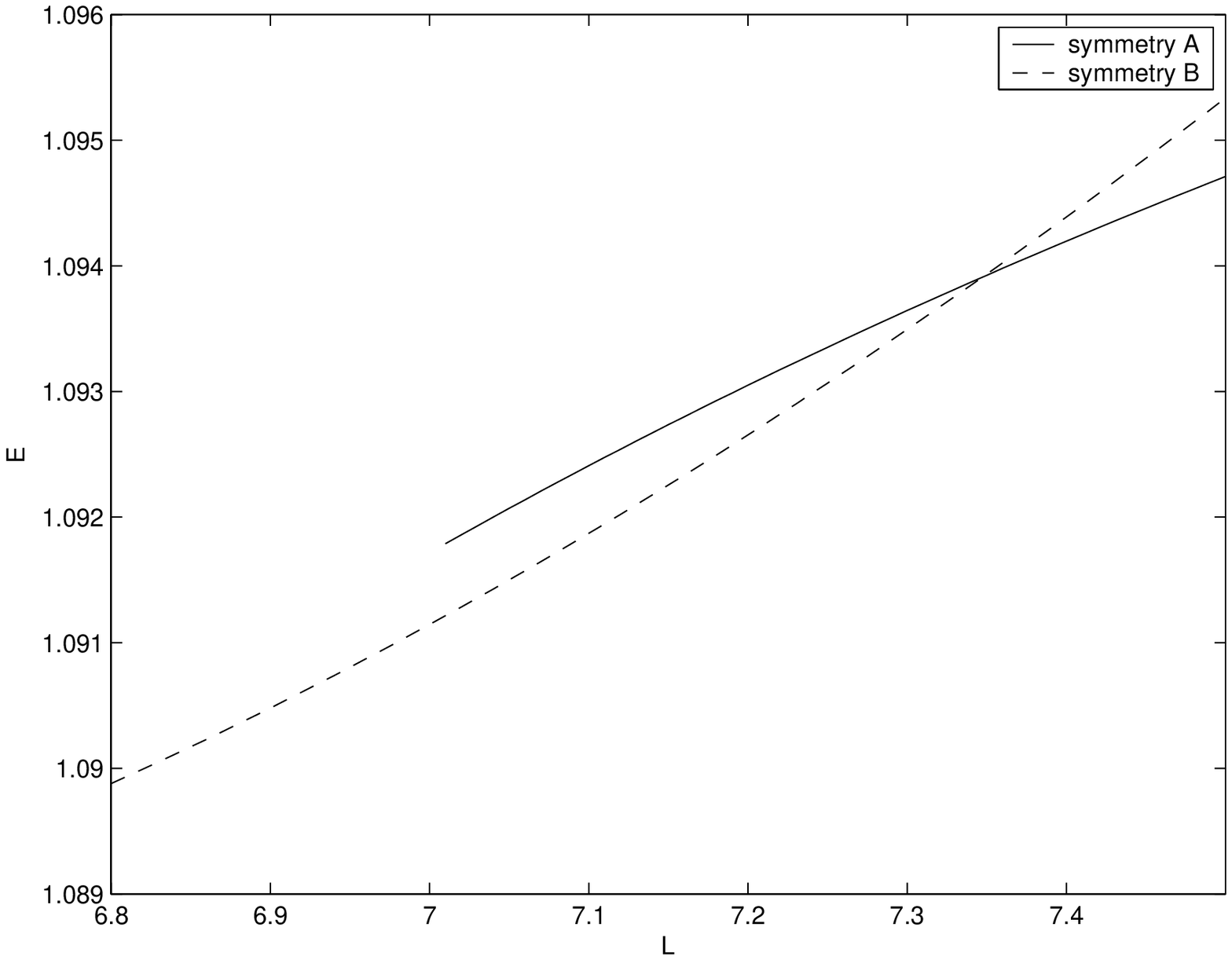}
}
\quad
\subfigure{
\includegraphics[scale=0.4]{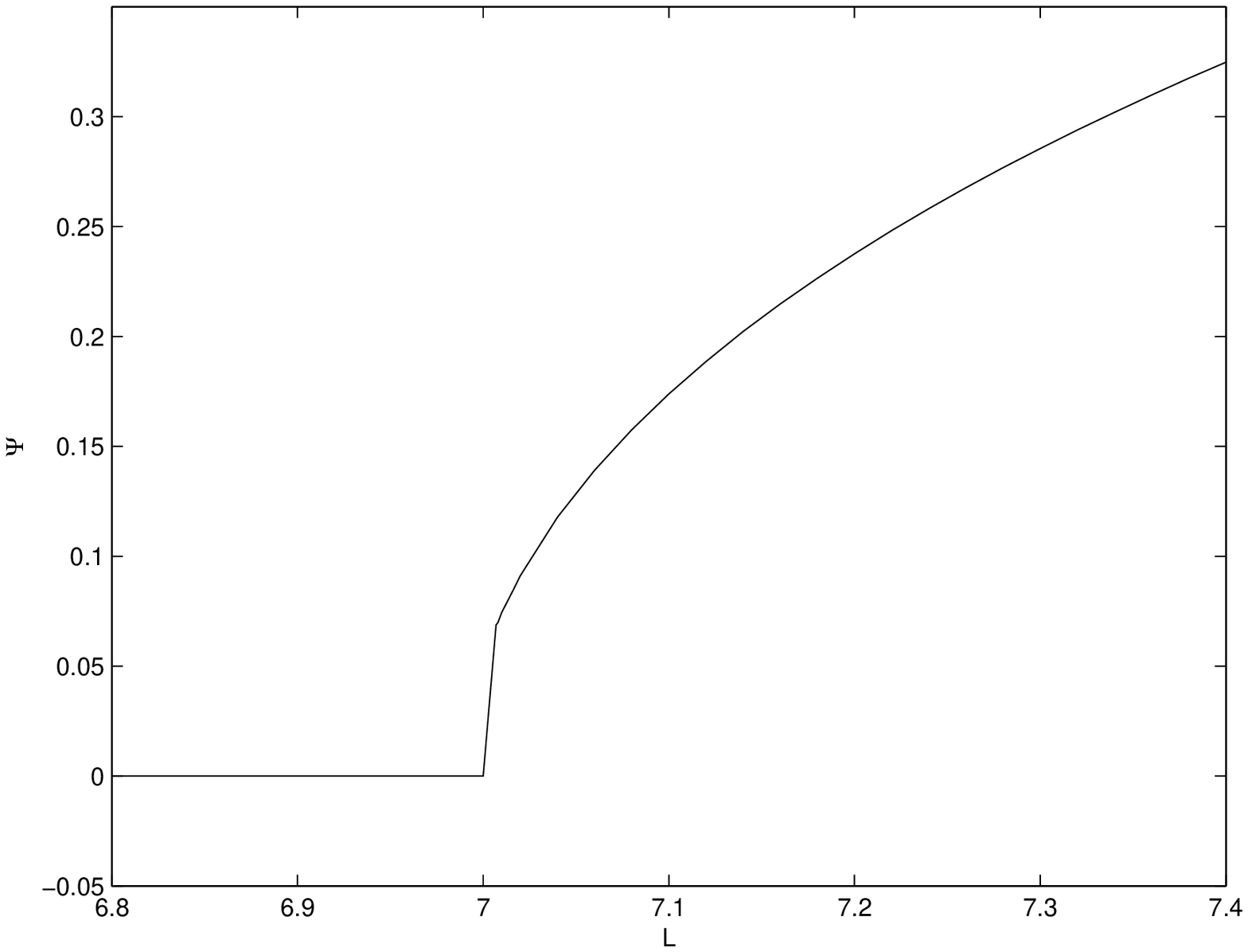}
}
\caption{$E$ and $\Psi=\langle\phi_3\rangle_A$ for the
  periodic $N=2$ solutions.  \label{fig1}}
}
\end{center}
\end{figure}

Let us now compare this to the situation on the torus $T^2$.
So we are looking for doubly-periodic solutions --- periodic in both
$x^1$ and $x^2$ with period $L$.  If $\alpha=0$ ({\em ie} for the O(3)
sigma-model), then solutions correspond to elliptic functions,
and these have topological charge $N=2$ in the unit
cell.  So we expect the minimal-energy crystal in the Skyrme case
also to have an $N=2$ cell.  Such solutions can be found by
numerical minimization of the energy functional.  It turns out
there are two different crystals, namely one with (approximate)
symmetry $A$:
\begin{description}
  \item[A1.] $(x,y)\mapsto(-x,y)$,
       $(\phi_1,\phi_2,\phi_3)\mapsto(\phi_1,-\phi_2,\phi_3)$,
  \item[A2.] $(x,y)\mapsto(y,x)$,
       $(\phi_1,\phi_2,\phi_3)\mapsto(-\phi_1,\phi_2,\phi_3)$,
  \item[A3.] $(x,y)\mapsto(x+L/2,y)$,
       $(\phi_1,\phi_2,\phi_3)\mapsto(\phi_3,-\phi_2,\phi_1)$;
\end{description}
and one with symmetry $B$:
\begin{description}
  \item[B1.] $(x,y)\mapsto(-x,y)$,
       $(\phi_1,\phi_2,\phi_3)\mapsto(\phi_1,-\phi_2,\phi_3)$,
  \item[B2.] $(x,y)\mapsto(y,x)$,
       $(\phi_1,\phi_2,\phi_3)\mapsto(-\phi_3,\phi_2,-\phi_1)$,
  \item[B3.] $(x,y)\mapsto(x+L/2,y)$,
       $(\phi_1,\phi_2,\phi_3)\mapsto(-\phi_1,-\phi_2,\phi_3)$.
\end{description}
In the $\alpha=0$ system, symmetries $A$ and $B$ are equivalent;
but this degeneracy is broken by the potential term $V$.  The
situation is illustrated in Figure \ref{fig1}, which plots the normalized
energy $E$ for the two solutions, and the quantity
$\Psi=\langle\phi_3\rangle_A$ for the type-$A$ solution, as functions
of the periodicity $L$.
The value of the parameter $\alpha$ used in obtaining these graphs
is $\alpha=0.1$.  The Bogomolny bound (\ref{Bogbound}) for this system is
$E\geq1+\pi\alpha/4=1.0785$.  The lowest crystal energy is
attained when $L=L_0=6.0$, {\em ie} at density $\rho=0.056$;
this solution has symmetry of type $B$, and normalized energy
$E_{{\rm crys}}=1.088$.  The type-$B$ solution exists for lower density,
and is always delocalized, {\em ie} has $\langle\phi_3\rangle=0$.

At relatively high density, in fact for $L<L_1=7.0$, the solution with type-$A$
symmetry also has $\langle\phi_3\rangle=0$; but it is unstable, and its energy
(which is not plotted in Figure \ref{fig1}) is higher than that of the
type-$B$ solution. When perturbed, it decays to the type-$B$ solution.
For $L>L_1$, the type-$A$ solution delocalizes ($\Psi\neq0$) and becomes stable.
(It no longer has the symmetry $A$ --- in fact $A3$ is invalid --- but we still
refer to it as the type-$A$ solution.)  This is illustrated in the right-hand
diagram of Figure \ref{fig1}; the numerical results indicate that
\[
   \lim_{L\to L_1+} \Psi \approx 0.06,
\]
but this limit is difficult to study owing to the instability which sets in
below $L_1$.
In the range $L_1<L<L_2=7.35$, the energy of the type-$A$ solution is
greater than that of
type $B$ --- in other words, there are two local mimina of the energy,
one of which (type $B$) is more symmetric than the other.  For $L>L_2$,
the minimal-energy $N=2$ solution is the localized type-$A$ one.

We see, therefore,  that there is a transition between a dense,
highly-symmetric homogeneous phase and a low-density, less homogeneous
phase.  The latter appears at  $L_1=7.0$, corresponding to a density
$\rho=2/7^2=0.041$, and it becomes the minimal-energy state at $L_2=7.35$,
corresponding to the critical density
\begin{equation}\label{NBScrit2}
  \rho_{{\rm c}}=2/7.35^2=0.037\,.
\end{equation}
The close agreement between (\ref{NBScrit2}) (for $N=2$ Skyrmions
on $T^2$) and (\ref{NBScrit1}) (for $N=1$ Skyrmions on $S^2$) is
remarkable.

Let us now consider the zero-density limit $L\to\infty$.  The only
known static solutions on $\RR^2$ are rotationally-symmetric rings \cite{We99}.
The energy of these goes like $E\approx\beta+\gamma/N^2$, where
$\beta$ and $\gamma$ are constants.  The significant fact here is that
$\beta$ is lower than $E_{{\rm crys}}$; this has been checked
numerically for $\alpha\leq1$.  For example, for $\alpha=0.1$, 
we get $\beta=1.082$, compared with $E_{{\rm crys}} = 1.088$.
The implication
of this is that, unlike the case for three-dimensional Skyrmions,
isolated crystal chunks are not energetically favourable, even for
large values of $N$.  In fact, it seems likely that
the minimum-energy configuration on $\RR^2$ is, for all values of $N$,
a single O(2)-symmetric ring (other local minima may, however, exist).
In this respect, the system is different from the three-dimensional
Skyrme model.


\section{The System $V =(1-\phi_3^2)(1-\phi_1^2)$}

The motivation for the potential $V$ studied in this section comes
from requiring that the Bogomolny bound (\ref{Bogbound}) be saturated
by a doubly-periodic solution.  As a consequence of this, the energy
of the corresponding Skyrme crystal will be as low as it possibly can be;
and so an isolated multi-Skyrmion will, for $N$ large enough, take the form
of a crystal chunk rather than a ring-like configuration.  In this
respect, it will be a better analogue of the three-dimensional Skyrme system.

Our assumption, therefore, is that the system should admit a solution
$W(z)$ which is an elliptic function.  One particularly simple choice
is to take $W(z)=2\wp(z)$, where $\wp$ is the Weierstrass p-function
with parameters $g_2=1$ and $g_3=0$.  So the function $F$
appearing in (\ref{Bogpot}) is given by $F(W)^2=2W(W^2-1)$; and the
corresponding potential function is easily seen to be
\begin{equation}\label{BBS}
   V(\vphi) = 16\,(1-\phi_3^2)(1-\phi_1^2).
\end{equation}
For the rest of the section we shall use this $V$.

The fundamental cell of the lattice is a square with sides of length
$L_0=2\,K(0.5)=3.708$, where $K(m)$ is the complete elliptic integral
of the first kind with parameter $m$.  This cell contains two units of
topological
charge (that is, the map from $T^2$ to $S^2$ has winding number $N=2$);
the distribution of charge- and energy-density indicates that it should
be thought of as a square lattice of half-Skyrmions.  It has the symmetry
of type $A$ defined in the previous section.
If we scale the solution to fit a lattice with edge-length $L$, then
we get a solution with normalized energy
\begin{equation}\label{Elat}
   E(L) = 1 + 2.3729\times\frac{\alpha}{2}
               \left(\frac{L}{L_0}+\frac{L_0}{L}\right).
\end{equation}
If $L=L_0$, the solution saturates the lower bound (\ref{Bogbound}), which
in this case is $E \geq 1 + 2.3729\,\alpha$.

For large $L$ (low density) one expects that this solution (a scaled
Weierstrass function) will no longer be the minimal-energy
doubly-periodic $N=2$ solution. In fact, the minimal
solution will correspond to an isolated $N=2$ lump at some point 
on $T^2$ --- and so will have $\langle\phi_3\rangle\neq0$.
\begin{figure}[htb]
\begin{center}
{
\includegraphics[scale=0.6]{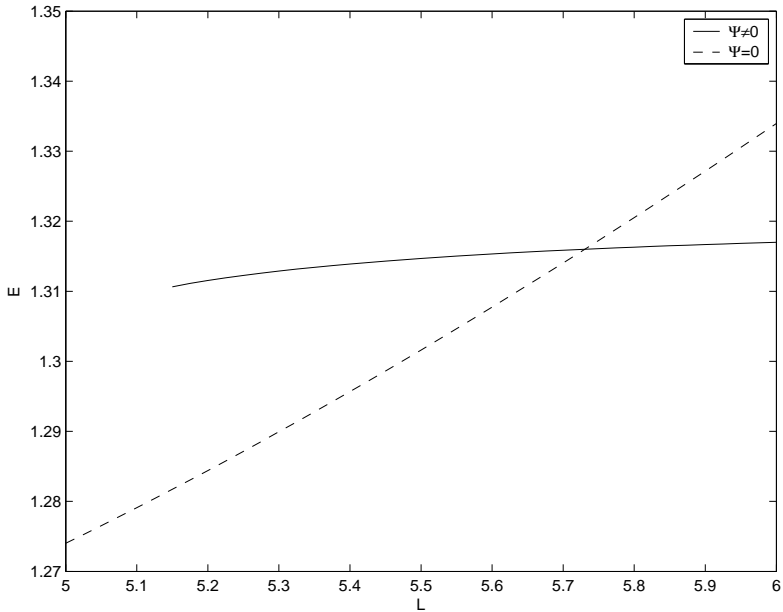}
\caption{Energy $E$ for the two $N=2$ solutions.  \label{fig2}}
}
\end{center}
\end{figure}
The result of a numerical
investigation of this phenomenon is presented in Figure \ref{fig2};
it was obtained by finding local minima of the energy with
$N=2$ for a range of values of the periodicity $L$. The value of $\alpha$
was taken to be $\alpha=0.1$.  For $L<5.15$, there is only one
solution, namely the Weierstrass-type one with $\langle\phi_3\rangle=0$.
Its energy grows fairly rapidly with $L$, but it remains a stable
solution ({\em ie} a local minimum), at least up to $L=20$. For $L\geq5.15$,
another solution (local minimum) appears --- a spatially-localized one with
$\langle\phi_3\rangle\neq0$.  Its energy grows less
rapidly with $L$, and for $L>5.73$ it is the lowest-energy solution.
In this sense, there is a phase transition at $L_1=5.73$, corresponding to
the critical density $\rho_{{\rm c}}=2/5.73^2=0.061$.
It would be interesting to study this transition in other ways, for example
by examining the system on a sphere (as in \cite{M87,IW01}) or by Monte-Carlo
simulations at finite temperature (as in \cite{SW02}).

\begin{figure}[htb]
\begin{center}
{
\subfigure{
\includegraphics[scale=0.4]{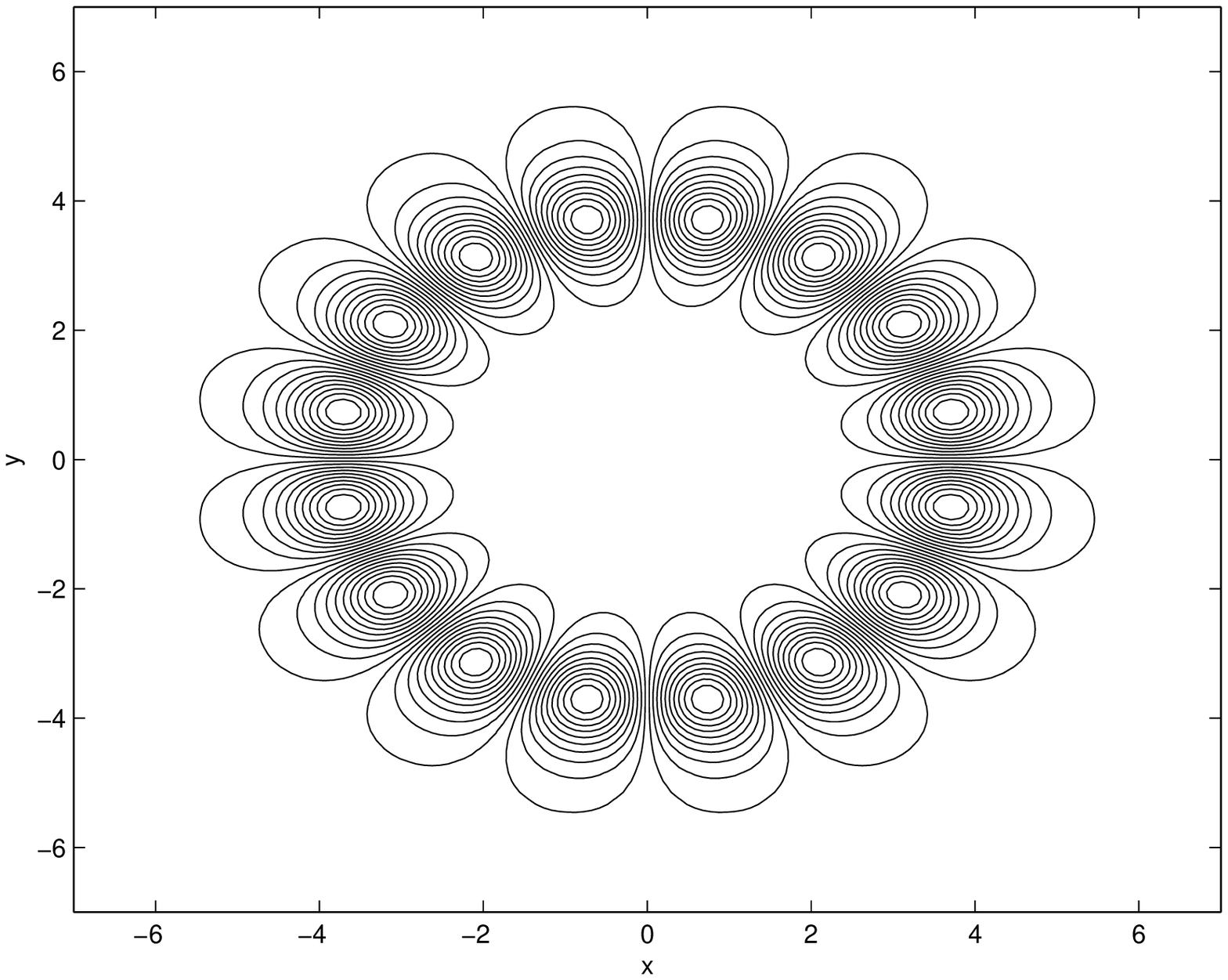}
}
\quad
\subfigure{
\includegraphics[scale=0.4]{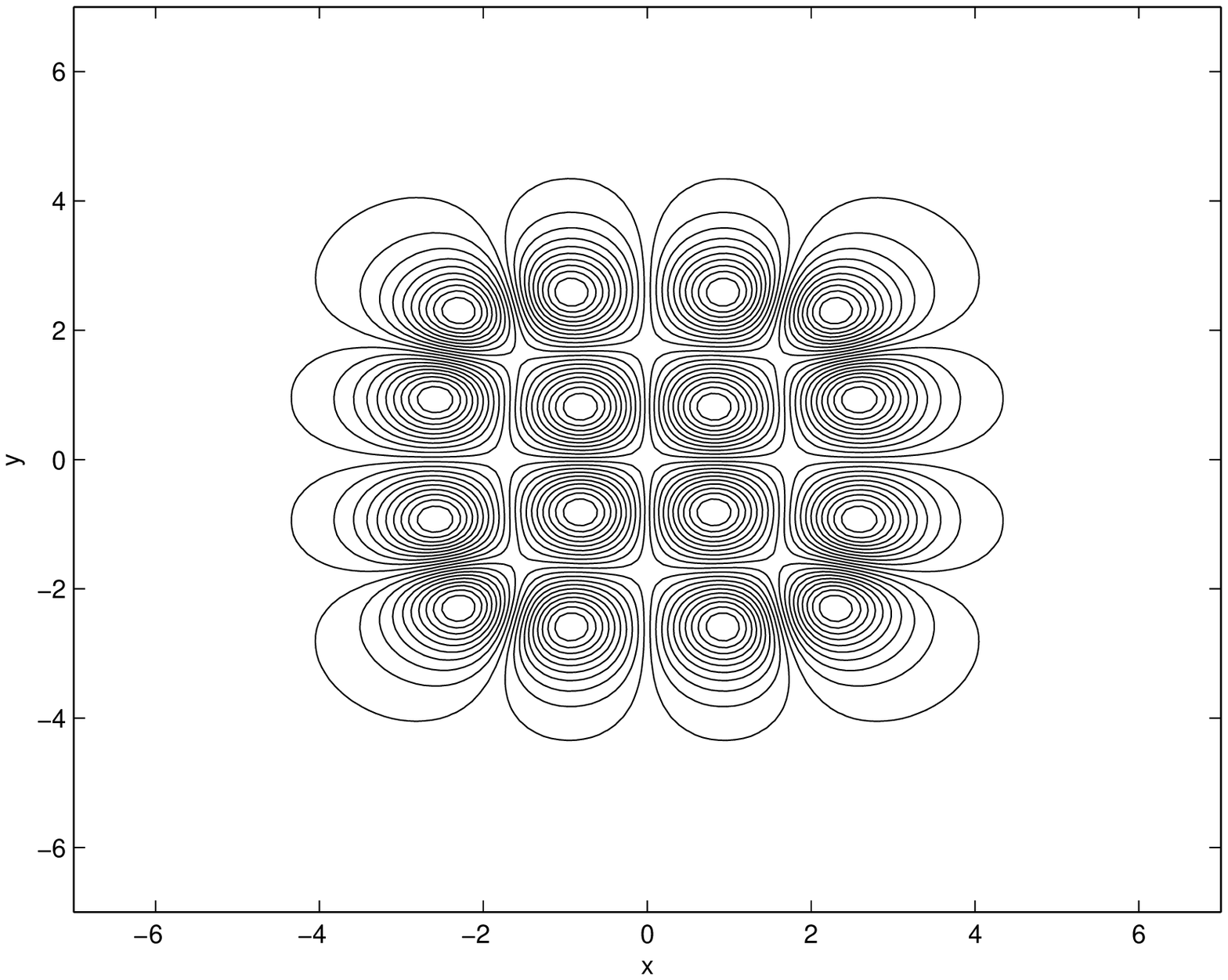}
}
\caption{Contour plots of $\phi_2$ for two static $N=8$ solutions
  of the system (\ref{BBS}).  \label{fig3}}
}
\end{center}
\end{figure}

Let us now look at the $L\to\infty$ limit, and consider finite-$N$
solutions on the infinite plane $\RR^2$, with the boundary condition
$\phi_3=1$ at spatial infinity.  First, the system admits ring-like solutions.
For these, we have $\phi_3=-1$ at a single point $O$, and
$\phi_3=0$ on a deformed ring centred at $O$.  Because of the factor
$(1-\phi_1^2)$ in (\ref{BBS}), this ring is not quite rotationally-symmetric
about $O$; its symmetry group is a subgroup of the rotation group O(2),
namely the dihedral group $D_{2N}$.  In other words, these solutions resemble
polygonal rings, analogous to the polygonal shells of the
three-dimensional Skyrme system.  A numerical investigation (for the
case $\alpha=0.1$) reveals that their normalized energy $E$ is very well
fitted by the function
\begin{equation}\label{Ering}
     E=1.262+0.227/N^2
\end{equation}
for $N\geq2$.  For example, the energy
of the $N=8$ ring is $E=1.266$; for comparison, the topological lower
bound is $E=1.2373$.  This $N=8$ ring is depicted in the left-hand plot of
Figure \ref{fig3}; it is a stable static solution with $D_{16}$
symmetry, and is best thought of as a ring of sixteen half-Skyrmions.

The other localized solution illustrated in Figure \ref{fig3}
corresponds to a clump of the Skyrme crystal --- in the right-hand plot,
we see a $2\times2$ clump of crystal, again containing sixteen
half-Skyrmions.  The energy of such a clump is approximately equal to
the (bulk) energy
of the crystal plus a surface contribution.  If the circumference of the
clump has length $ML_0$, where $M$ is a positive integer, then one expects
that the normalized energy will have the form
\begin{equation}\label{Eclump}
   E \approx E_0 + M\,\delta E\,/N
\end{equation}
where $E_0 = 1 + 2.373\,\alpha$ and
where $\delta E$ is the contribution from a single edge of lattice-cell.
One may estimate $\delta E$ as follows (this is analogous to the discussion
in \cite{B96}).  Start with a single cell, with coordinate range
$0\leq x\leq L_0$, $0\leq y\leq L_0$.  Now stretch the cell in the
positive $x$-direction, so that the $x$-range becomes $0\leq x<\infty$.
The boundary conditions are that the field $\vphi$ remains periodic in the
$y$-direction, maintains its original lattice-configuration on the
edge $x=0$, and tends to its asymptotic value $(0,0,1)$ as $x\to\infty$.
We can then relax this configuration, minimizing its energy numerically
to obtain a mimimum energy $E_1$.  The deformed cell still contains
two units of Skyrme charge, so the resulting estimate for $\delta E$ is
\begin{equation}\label{deltaE}
   \delta E \approx E_1 - 2E_0.
\end{equation}
For example, if $\alpha=0.1$, one gets $\delta E=0.026$.
Consequently, our estimate for the energy of a localized square
lattice clump of charge $N$, with $\alpha=0.1$, is
\begin{equation}\label{Eclumpp}
   E \approx E_0 + 2\sqrt{2}\,\delta E/\sqrt{N} = 1.237+0.074/\sqrt{N};
\end{equation}
but for low $N$ this will be an underestimate, since there are also corner
effects to be considered.  These corner effects presumably give an additional
contribution to $E$ of the form $C/N$, where $C$ is a constant.

If we take a single lattice cell ($N=2$), then it resembles a square
array of four half-Skyrmions, but this amounts to the same thing as a
ring of four half-Skyrmions --- indeed, the $N=2$ solution on $\RR^2$ has
a unique shape.  For $N=8$, however, there is the possibility of the
two  solutions depicted in Figure \ref{fig3}, and each of these is in
fact realized, as a local minimum of the energy.
The formulas (\ref{Eclumpp}) and (\ref{Ering}) suggest that the energy
of the $N=8$ ring should be higher than that of the square clump, by an
amount $\Delta E = 0.0027$. The numerical result is that $\Delta E = -0.0035$,
which gives an estimate for the corner contribution which has to be added
to (\ref{Eclumpp}), namely $0.05/N$.

For larger crystal chunks, the ring will be less favourable.  For example,
we expect, from (\ref{Ering}) and the modified version of (\ref{Eclumpp}),
that the minimal-energy $N=18$ solution will correspond to a $6\times6$ square
array of half-Skyrmions; and in particular, that the energy of the $N=18$
ring will be higher than that of the $6\times6$ clump by an amount
$\Delta E = 0.0057$.  It should be emphasized, however, that there are
likely to be many local minima, of which the square chunk and the ring are
just two examples.  All this is very similar to the situation for
the three-dimensional Skyrme system.

In conclusion, this particular planar system, in addition to admitting an
explicit crystal solution which saturates the topological lower bound on
the energy, is a good model for three-dimensional Skyrmions.




\end{document}